\pdfoutput=1
\documentclass[conference]{IEEEtran}
\IEEEoverridecommandlockouts{}
\usepackage{cite}
\usepackage{amsmath,amssymb,amsfonts}
\usepackage{algorithmic}
\usepackage{graphicx}
\usepackage{textcomp}
\usepackage{xcolor}
\usepackage[inline]{enumitem}

\def\baselinestretch{0.96} 

\def\BibTeX{{\rm B\kern-.05em{\sc i\kern-.025em b}\kern-.08em
    T\kern-.1667em\lower.7ex\hbox{E}\kern-.125emX}}

\usepackage[]{algorithm2e}

\DeclareMathAlphabet{\pazocal}{OMS}{zplm}{m}{n}
\SetMathAlphabet\pazocal{bold}{OMS}{zplm}{bx}{n}

\newcommand*{\xhat}[1]{#1\kern-0.53em\hat{\phantom{#1}}}
\newcommand*{\xzhat}[1]{#1\kern-0.7em\hat{\phantom{#1}}}
\newcommand*{\xxhat}[1]{#1\kern-1.0em\hat{\phantom{#1}}}
\newcommand*{\xtilde}[1]{#1\kern-0.53em\tilde{\phantom{#1}}}
\DeclareMathOperator*{\argminA}{argmin \hspace{1.1mm}}

\newcommand{\bbC}{\ensuremath{\mathbb{C}}}

\newcommand{\bx}{\ensuremath{\mathbf{x}}}

\newcommand{\bn}{\ensuremath{\mathbf{n}}}

\newcommand{\bs}{\ensuremath{\mathbf{s}}}

\newcommand{\by}{\ensuremath{\mathbf{y}}}

\newcommand{\balpha}{\ensuremath{\boldsymbol{\alpha}}}
\newcommand{\btheta}{\ensuremath{\boldsymbol{\theta}}}

\newcommand{\bphi}{\ensuremath{\boldsymbol{\phi}}}

\newcommand{\bPhi}{\ensuremath{\boldsymbol{\Phi}}}
\newcommand{\bTheta}{\ensuremath{\boldsymbol{\Theta}}}

\newcommand{\bX}{\ensuremath{\mathbf{X}}}

\newcommand{\bM}{\ensuremath{\mathbf{M}}}
\newcommand{\bP}{\ensuremath{\mathbf{P}}}
\newcommand{\bH}{\ensuremath{\mathbf{H}}}

\newcommand{\bA}{\ensuremath{\mathbf{A}}}

\newcommand{\bI}{\ensuremath{\mathbf{I}}}

\newcommand{\bN}{\ensuremath{\mathbf{N}}}
\newcommand{\bT}{\ensuremath{\mathbf{T}}}
\newcommand{\bU}{\ensuremath{\mathbf{U}}}

\newcommand{\tr}{\ensuremath{\text{Tr}}}

\SetKwInput{KwInput}{Input}
\SetKwInput{KwOutput}{Output}
\SetKwFunction{FHank}{hankel}
\SetKwFunction{FCons}{construct}
\SetKwFunction{FEsimateWeig}{estWeight}
\SetKwFunction{FWeigSubFit}{solNLS}
\SetKwFunction{FLS}{solLS}

\SetKwFunction{SVD}{SVD}
\SetKwFunction{EVD}{EVD}
\SetKwFunction{TSVD}{TSVD}
\SetKwFunction{JDIAG}{JDIAG}
\SetKwFunction{MPS}{calPseudoSpectrum}
\SetKwFunction{SP}{searchPeaks}
\SetKwFunction{find}{find}
\SetKwFunction{rowstack}{rowstack}
\SetKwFunction{colstack}{colstack}

\SetKwFunction{unvec}{unvec}
\SetKwFunction{EVD}{EVD}

\usepackage{pifont}

\begin{document}

\title{Joint Multiple FMCW Chirp Sequence Processing for Velocity Estimation and Ambiguity Resolving \\}

\author{\IEEEauthorblockN{Tarik Kazaz,
		Karan Jayachandra,  Arie Koppellar and
		Yiting Lu}
	\IEEEauthorblockA{NXP Semiconductors,
		Eindhoven, The Netherlands\\
		Email: \{tarik.kazaz, karan.jayachandra, arie.koppellar,
		yiting.lu\}@nxp.com}
}

\maketitle

\begin{abstract}
In FMCW automotive radar applications, it is often a challenge to design a chirp sequence that satisfies the requirements set by practical driving scenarios and simultaneously enables high range resolution, large maximum range, and unambiguous velocity estimation. To support long-range scenarios the chirps should have a sufficiently long duration compared to their bandwidth. At the same time, the long chirps result in ambiguous velocity estimation for targets with high velocity. The problem of velocity ambiguity is often solved by using multiple chirp sequences with co-prime delay shifts between them. However, coherent processing of multiple chirp sequences is not possible using classical spectral estimation techniques based on Fast Fourier Transform (FFT). This results in statistically not efficient velocity estimation and loss of processing gain. In this work, we propose an algorithm that can jointly process multiple chirp sequences and resolve possible ambiguities present in the velocities estimates. The resulting algorithm is statistically efficient, and gridless. Furthermore, it increases the resolution of velocity estimation beyond the natural resolution due to its super-resolution properties. These results are confirmed by both numerical simulations and experiments with automotive radar IC.
\end{abstract}


\section{Introduction}
Automotive radars together with other sensors such as lidar, camera, and ultrasound are key components of self-driving vehicles and advanced driver-assistance systems (ADAS) \cite{patole2017automotive}. In these applications, it is of interest to infer information about the surroundings of the vehicles and based on it assist the driver in controlling the vehicles or fully autonomously control the vehicle. In these sensing problems, the key challenge is to achieve high-resolution estimation of objects surrounding the vehicle in various parameter domains. In the context of signal processing, these problems belong to the group of high-resolution source estimation and localization and they arise in diverse application areas such as medical imaging \cite{8878141}, radio astronomy \cite{4291873}, RF localization \cite{kazaz2022multiband}, radar \cite{herman2009high}, and sonar \cite{trucco2008devising} sensing, to name a few. In this paper we focus on radar sensor where the goal is to provide information about ranges, velocities, azimuth, and elevation of the objects that are surrounding the vehicle. These objects in radar terminology are called targets. The popular approach for the implementation of automotive radar sensors is based on multiple-input multiple-output (MIMO) coding and frequency-modulated continuous wave (FMCW) modulation \cite{engels2017advances}. The MIMO coding is used to increase the angular resolution of the radar systems without introducing additional physical radio frequency (RF) chains. On the other hand, FMCW modulation support stretch processing and thus enables cheap radar sensor implementation due to the low sampling rate and low peak-to-average ratio (PAPR) requirements.

However, implementation of automotive MIMO FMCW radar sensors is challenging when system requirements set by the automotive use cases are considered. For example, common automotive radar use cases require a large field-of-view (FOV) and maximum detectable range, unambiguous velocity estimation, and high angular resolution. Some of these requirements are contrary to each other and often require design trade-offs to be made to satisfy them simultaneously \cite{sun2020mimo}. In particular,  it is challenging to increase the maximum range and maximum unambiguous velocity at the same time with the single FMCW chirp sequence. This problem becomes even more difficult when Doppler Division Multiplexing (DDM) is used as a MIMO scheme for multiplexing transmit (Tx) antennas as it further reduces the maximum unambiguous velocity estimation \cite{sun2022enhancing}. The DDM is a popular MIMO scheme that enables transmission from multiple Tx antennas at the same time increasing the power budget and FOV compared to the scenario when the time division multiple access (TDMA) scheme is used \cite{jansen2019automotive}. Therefore, to support DDM and satisfy both requirements on the maximum detectable range and unambiguous velocity, typically multiple chirp sequences are used where their pulse repetition interval (PRI) and the relative time delays between sequences are carefully designed such that they create co-prime sampling frequencies in the slow-time domain. This co-primality property of the chirp sequences is then used to resolve the velocity ambiguities \cite{kronauge2010radar}.

However, coherent processing of multiple chirp sequences is not possible with FFT based algorithm resulting in a loss of algorithmic processing gain. In this paper, we focus on the problem of joint coherent processing of multiple chirp sequences in MIMO FMCW radar. Our aim is to develop a statistically efficient algorithm that enables joint velocity estimation and ambiguity resolving from all available samples and chirps sequences. In particular, by constructing the block Hankel matrix from the samples of multiple chirp sequences the multiple shift-invariance structure \cite{swindlehurst1992multiple} of the data model is reviled \cite{237536}. A similar data model appears in multiband ranging \cite{9555252} and array signal processing \cite{viberg1991sensor} when antenna subarrays with multiple apertures are used. Therefore, the related algorithms to this model are applicable. We use this property of the data to design an algorithm based on subspace estimation \cite{237536} and nonlinear least squares minimization \cite{golub2003separable} to jointly estimate the velocities of the targets and their ambiguities. 

The resulting algorithm is benchmarked through simulations and real experiments with NXP SAF85xx radar IC \cite{nxp}, by comparing its performance against the algorithm for velocity estimation and ambiguity resolution based on FFT. The results show that the proposed algorithm enables coherent processing of multiple chirp sequences, removes grid bias, increases the resolution of velocity estimation, and provides approximately 8 dB of SNR gain compared to the classical FFT approach.
\section{Problem Formulation and System Modeling}
\begin{figure}[h!]
	\centering
	\includegraphics[width=0.45\textwidth]{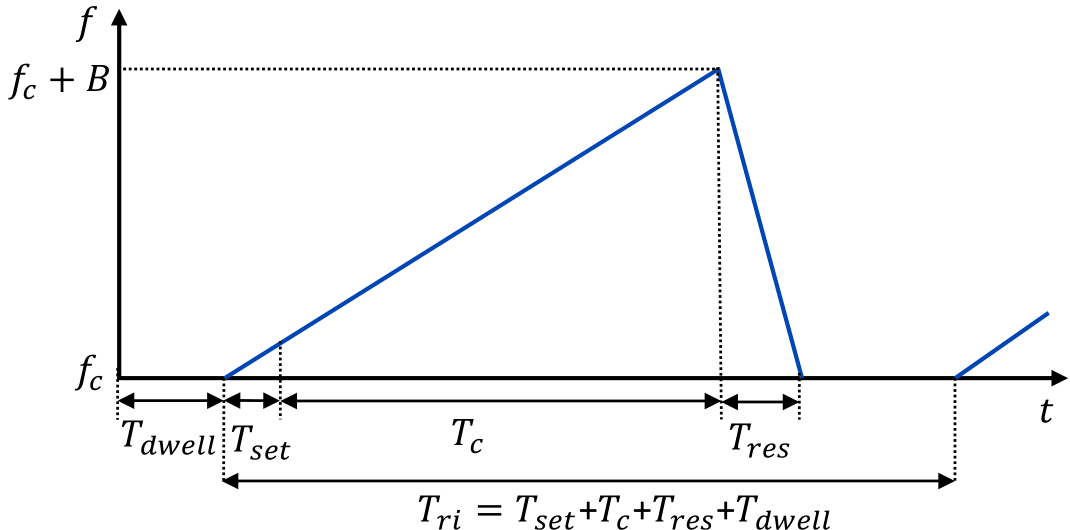}
	\caption{Illustration of FMCW chirp with its bandwith $B$, dwell $T_{dwell}$, set $T_{set}$, reset $T_{res}$, acquisition $T_c$ and repetition interval $T_{ri}$ time parameters.} 
	\label{fig:chirp}
\end{figure}

\begin{figure}
	\centering
	\includegraphics[width=0.5\textwidth]{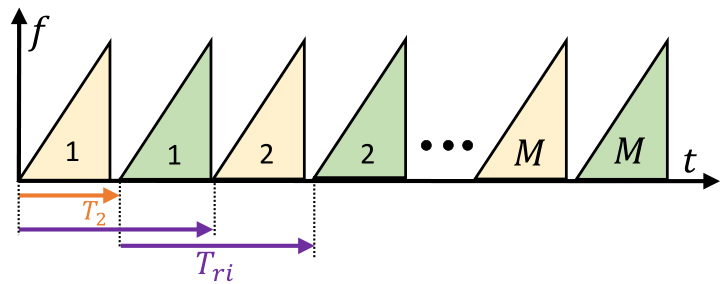}%
	\caption{Illustration of two FMCW sequences with $M$ chirps, the relative time shift $T_{2}$ and the same chirp repetition interval $T_{ri}$.} 
	\label{fig:fmcw_multiple_sequence}
\end{figure}

In this section, we discuss the assumptions on the radar scenario and derive the corresponding signal model. We consider MIMO FMCW radar that has $K_{Tx}$ Tx antennas and uses DDM scheme to orthogonalize transmitted signals. The transmitters send $M$ FMCW chirps during the single measurement sequence. These chirps reflect from the targets and superimpose at the arrival at $K_{Rx}$  receive (Rx) antennas. The receiver de-chirps the reflected signals and performs sampling of the continuous-time beat signal. During the single chirp acquisition interval $T_c$ (cf. Figure \ref{fig:chirp}), $N$ samples of the beat signal are collected, that is $T_c = N T_s$, where $T_s =  1/f_s$ is the sampling period. Therefore, during the single full chirp sequence (cf. Figure \ref{fig:fmcw_multiple_sequence}), $M$ sets of $N$ fast-time samples are collected. 

Given the parameters of the chirp sequences, we can define the maximum unambiguous velocity, maximum range, and range resolution that can be estimated from it. Assuming the real sampling of the beat signal the range resolution $\Delta R$ and maximum range $R_{\max}$ are given by

\begin{equation}
	\label{eq:max_res_range}
	\Delta R = \frac{c}{2B} \,, \qquad  \quad R_{\max} = \frac{c f_s T_c}{4B} \,,
\end{equation}

\noindent where  $B$ is the bandwidth of the chirp, $\lambda = c/f_c$ is the wavelength of the signal, $c$ is the speed of light in the air and $f_c$ is the carrier frequency of the signal. The maximum absolute velocity and Doppler frequency of the targets that can be unambiguously estimated using a single chirp sequence are given by
\begin{equation}
	\label{eq:un_vel_mimo}
	\lvert v_{\max} \rvert < \frac{\lambda}{4 K_{Tx} T_{ri}}  \iff \lvert f_{d, \max} \rvert < \frac{1}{2 K_{Tx} T_{ri}}\,.
\end{equation}
\noindent From the equations (\ref{eq:max_res_range}) and (\ref{eq:un_vel_mimo}) we can observe that to lower $\Delta R$ the bandwidth of the chirps $B$ shall be increased. However, an increase in bandwidth at the same time decreases $R_{\max}$. The maximum range could be increased by increasing the chirp acquisition interval $T_c$. Still, from Fig. \ref{fig:chirp} we can observe that increasing the $T_c$ would increase the $T_{ri}$ reducing the maximum unambiguous velocity. Therefore, these requirements are conflicting to each other which motivates the use of multiple chirp sequences to satisfy these requirements in practical automotive scenarios. 

\subsection{Signal Model}
We assume that for the purpose of velocity ambiguity resolving radar sends $L$ full chirp sequences with $M$ chirps during the one radar detection cycle. Without loss of generality, to simplify further exposition we will consider the signal model for samples collected at one of the Rx antennas. The signal model for the samples collected during a single $l$th chirp sequence reflected from $P$ targets can be written as
\begin{equation}
	\label{eq:model1}
	\begin{aligned}
		\tilde{s}(n, m, l) = \sum_{k=1}^{K_{Tx}} \sum_{p=1}^{P} & \alpha_p \exp\bigg\{j2\pi \bigg[\left(\frac{2\eta R_p}{c} + f_{d,p}\right) \frac{n}{f_s} \\
		& + (f_{d,p} + f_{ddm, k}) mT_{ri} + f_{d,p}T_l \\
		& + \frac{f_c d_k \text{sin}(\beta_p)}{c} \bigg] \bigg\}\,,
	\end{aligned}
\end{equation}

\noindent where range, Doppler frequency, and direction-of-departure (DOD) of the $p$th target are denoted by $R_p$, $f_{d,p}$, and $\beta_p$ parameters, respectively. The velocity of the target directly follows from Doppler frequency as $v_{d,p} = f_{d,p} / 2c$. The slope of the chirp is denoted by $\eta$, $f_c$ is the central frequency of the chirp, $f_{ddm, k}$ and $d_k$ are DDM frequency and azimuth position of $k$th transmitter. The indices of fast and slow-time samples are denoted by $n = 1,\dots, N$, and $m = 1,\dots, M$, respectively. The time shift of $l$th chirp sequence compared to the reference sequence is denoted by $T_l$.

In radar applications, the goal is to estimate parameters of the targets $R_p$ $\alpha_p$, $f_{d,p}$, and $\beta_p$, $p=1, \dots, P$ from the collected samples (\ref{eq:model1}). In this paper, we focus on the problem of coherent processing of samples from multiple chirp sequences for velocity estimation and ambiguity resolving. That is we focus on estimation of parameters $f_{d,p}$, $p=1, \dots, P$. In what follows we derive and discuss the data model for the collected samples. These results we later use to derive the algorithm for joint Doppler estimation and ambiguity resolving (J-DEAR).
\section{Data Model}
The model for $n$th range frequency bin after windowing and range FFT are applied on the beat signal (\ref{eq:model1}), can be written as

\begin{equation}
	\begin{aligned}
		\label{eq:model2}
		s(\omega_n, m, l) &= \sum_{k=1}^{K_{Tx}} \sum_{p=1}^{P} \alpha_p W(\omega_n - \omega_{R_p}) \\ 
		& \exp\bigg\{j2\pi \bigg[ (f_{d,p} + f_{ddm, k}) mT_{ri} + f_{d,p}T_l \\
		& + \frac{f_c d_k \text{sin}(\beta_p)}{c} \bigg] \bigg\}\,,
	\end{aligned}
\end{equation}

\noindent where $\omega_n$ is $n$th bin discrete angular frequency, $W(\omega)$ is FFT of the window function that is applied on fast-time samples, and $\omega_{R_p} = 2\pi \left(\frac{2\eta R_p}{c} + f_{d,p}\right) \frac{n}{N}$. In this model, we do small abuse of notation and absorb normalization factor of FFT transform in $\alpha_p$, $p=1, \dots, P$ parameters as it affects all the targets equally.

From the model (\ref{eq:model2}) it can be observed that DDM frequencies of Tx's modulate each target $K_{Tx}$ times in Doppler domain. Therefore, in Doppler domain instead of $P$ targets, there are $P K_{Tx}$ targets present in the model due to the modulation replicas. However, the $K_{Tx}$ replicas of the same target are deterministically related to each other as their modulating frequencies $f_{ddm, k}$, $k=1, \dots, K_{Tx}$ are known at Rx. Therefore, the effects of DDM modulation can be taken into account when computing the final velocity estimate of each target from the velocity estimates of $K_{Tx}$ replicas as described in Section \ref{sec:algo}. Let us fix the range index number $n$  and avoid writing it to simplify notation as the model for every of the range bins will be structurally the same. Then we can write model (\ref{eq:model2}) for $m=1,\dots, M$ slow-time samples of $l$th chirp sequence in matrix form as
\begin{equation}
	\label{eq:model_m1}
	\bs_l = \bM\bTheta_l\bx + \bn \,,
\end{equation} 
where  $\bM \in \bbC^{M\times PK_{Tx}}$ is a block Vandermonde matrix which satisfies the model 
\begin{equation}
	\label{eq:f_matrix}
	\bM = \left[ \bM_1 \quad \bM_2 \quad \cdots \quad \bM_{K_{Tx}}\right] \,,
\end{equation}
\noindent $\bM_k$, $k = 1, \dots, K_{Tx}$ are also Vandermonde matrices 
\begin{equation}
	\label{eq:f_matrix2}
	\bM_k = \begin{bmatrix}
		1 & 1 & \cdots & 1 \\
		\Phi_{k, 1} & \Phi_{k, 2} & \cdots & \Phi_{k, P}\\
		\vdots & \vdots & \ddots & \vdots \\ 
		\Phi_{k, 1}^{{M-1}} & \Phi_{k, 2}^{M-1} & \cdots & \Phi_{k, P}^{M-1}
	\end{bmatrix} \,,
\end{equation}
\noindent and $ \Phi_{k, p} = e^{-j\phi_{k, p}}$, where $\phi_{k, p} = 2 \pi (f_{d, p} + f_{k, p}) T_{ri}$ is phase shift introduced by DDM frequency of $k$th transmitter and velocity of the $p$th target over time aperture formed by two consecutive chirps within the same chirp sequence. The phase shifts introduced by Doppler frequency over time aperture between chirps in reference (i.e., first) and $l$th sequence are modeled by a diagonal matrix

\begin{equation}
	\label{eq:f_matrix3}
	\bTheta_l = \bI_{K_{Tx}} \otimes \text{diag}(\btheta_l)\,,
\end{equation}

\noindent where $\bI_{K_{Tx}}$ is $K_{Tx} \times K_{Tx}$ identity matrix, $\otimes$ denotes Kronecker product of two matrices, $\btheta_l = [\theta_{l,1},\dots,\theta_{l,P}]^T \in \bbC^P$, and $\theta_{l,p} = e^{j2 \pi f_{d, p} T_l}$. We can further write $\theta_{l,p} = \Phi_{k, p} e^{j2 \pi f_{k, p} T_l}$, which we will use in Section \ref{sec:algo} to parameterize data model and formulate velocity estimation problem. The power of the targets are modeled by $\bx = [(\balpha \odot \by_{DOD,1})^T, (\balpha \odot \by_{DOD,2})^T, \dots, (\balpha \odot \by_{DOD,K_{Tx}})^T]^T$, where $\balpha$ are complex amplitudes of the targets, $\by_{DOD,k} = [\gamma_{1, k}, \dots, \gamma_{P, k}]^T$, $\gamma_{1, k} = e^{j2\pi {f_c d_k \text{sin}(\beta_p)} / c}$ is phase shift introduced by DOD of $p$th target when illuminated by $k$th transmitter, $\odot$ is point-wise Hadamard product and $(\cdot)^T$ denotes transpose of an matrix. Lastly, vector $\bn$ represents the noise that corrupts the beat signal samples.

\section{Joint Multiple Chirp Sequence Processing}
\label{sec:algo}
Observing the model (\ref{eq:model_m1}) we can notice that slow-time samples collected using multiple chirp sequences have multiple shift invariance structures \cite{viberg1991sensor}. That is samples collected in different chirp sequences share the same column space up to phase shifts $\Theta_l$ introduced by velocities of the targets. In what follows, we use this property to develop an algorithm for joint velocity estimation and ambiguity resolving from multiple chirp sequences.

\subsection{Velocity Estimation}
Given the range bin vectors $\bs_l$, $l = 1, \dots, L$ the problem of velocity estimation and ambiguity resolving of the targets is to determine true parameters $f_{d,p}$, $p = 1, \dots, P$  without ambiguities. From the signal model (\ref{eq:model_m1}) it follows that parameters $f_{d,p}$ can be estimated from the column span of $\bM \Theta_l$. However, this signal subspace can not be directly estimated from the range bin vectors. To restore the rank, we construct Hankel matrices of size $B \times Q$ from these vectors as 
\begin{equation}
	\label{eq:hankel}
	\bH_l = \begin{bmatrix}
		s_l[0] & s_l[1] & \cdots & s_l[Q] \\
		s_l[1] & s_l[2] & \cdots &   s_l[Q+1] \\
		\vdots &  \vdots   &   \ddots     & \vdots  \\
		s_l[B-1] & s_l[B] & \cdots & s_l[M-1]
	\end{bmatrix} \,.
\end{equation}
Here, $B = M-Q-1$, and we require $B > P$ and $Q \ge P$. These parameter in practice can be tuned based on maximum expected number of targets that can be detected.  From (\ref{eq:model_m1}), and using the shift invariance of the Vandermonde matrix (\ref{eq:f_matrix}), the constructed matrices satisfy
\begin{equation} 
	\label{eq:hen_mat}
	\bH_l = \bM' \bTheta_{1l} \bX + \bN_l \,,
\end{equation}
\noindent where $\bM'$ is an $B \times (PK_{Tx})$ sub-matrix of $\bM$, $\bTheta_{1l}$ is diagonal matrix modeling phase shifts introduced by velocities of the targets between the first and $l$th chirp sequences, and $\boldsymbol{N}_l$ is a noise matrix for $l$th sequence.  Furthermore, \[\bX = [\bx,\, \bPhi \bx,\, \bPhi^2 \bx, \cdots, \bPhi^{Q-1} \bx] \,\] where $\mathbf{\Phi} = \text{diag}([\Phi_{1,1} \cdots \Phi_{K_{Tx},P}])$.	

To capture all the shift invariance structure present in the constructed Hankel matrices we stack them in block Hankel matrix as
\begin{equation}
	\label{eq:big_hank} 
	\nonumber 
	\bH =
	\begin{bmatrix} \bH_1 \\
		\bH_2 \\
		\vdots \\
		\bH_L
	\end{bmatrix} := 
	\begin{bmatrix} 
		\bM' \\
		\bM' \bTheta_{12} \\
		\vdots \\
		\bM' \bTheta_{1L}
	\end{bmatrix}
	\bX  + \bN\ = 
	\bA (\bPhi) \bX + \bN\,.
\end{equation}
\noindent where $\bTheta_{1l} = \bTheta_1^{-1}\bTheta_l$ and $(\cdot)^{-1}$ is inverse of an matrix. To parameterize $\bA(\bPhi)$ we have used property given in (\ref{eq:f_matrix3}) that $\bTheta_{1l}$, $l=1, \dots, L$ can be parameterized by $\bPhi$.

Assuming that the requirements on the dimension of the matrices (\ref{eq:hankel}) are satisfied and if all factors in (\ref{eq:big_hank}) are full rank, then $\bH$ has rank $PK_{Tx}$. Following this, we can estimate matrix $\bA$ from the column span of $\bH$ up to a $PK_{Tx} \times PK_{Tx}$ non-singular matrix $\bT$. Therefore, we can write ${\bA = \bU\bT^{-1}}$, where the columns of $\bU$ are the $PK_{Tx}$-dimensional basis vectors that span the column space of $\bH$.The matrix $\bU$ can be estimated by using singular value decomposition (SVD) from $\bH$. The signal subspace is then estimated by selecting left columns of $\bU$ that correspond to the $PK_{Tx}$ largest singular values. Typically number of largest singular values is not known and it can be detected by using Akaike information criterion (AIC) \cite{van2004optimum} or minimum description length (MDL) criteria \cite{bazzi2016detection}.

Since estimated column space preserves shift invariance structure introduced by the velocity of the targets we can estimate velocities by first estimating phase shifts $\bPhi$ from $\bU$. Taking into account the errors introduced during subspace estimation, we can write $\bA'(\bphi) \approx \hat{\bU}\bT^{-1}$. Following this property, we can estimate $\bPhi$ by using subspace estimation techniques. In particular, in this work, we formulate the following optimization problem to estimate $\bPhi$

\begin{equation} 
	\label{eq:sub_fit_prob1}
	\centering
	\hat{\bPhi}, \hat{\bT} = \argminA_{\bPhi, \bT} \left\|
	\hat{\bU} - \bA(\bPhi) \bT \right\|_{F}^2\,.
\end{equation}

\noindent This problem is a nonlinear least squares (NLS) problem that often arises in array signal processing \cite{viberg1991sensor, 9555252}. To further simplify it we will assume LS solution on $\bT$. That is, we assume that for the optimal $\bPhi$, the optimal $\bT$ must satisfy ${\bT = \bA^{\dagger}(\bPhi)\hat{\bU}}$. Based on this assumption, this problem can be further reformulated into the following separable nonlinear LS (SNLS) problem \cite{golub2003separable},
\begin{equation} 
	\label{eq:sub_fit_est1}
	\hat{\bPhi} = \argminA_{\bPhi}  \tr\left(\bP_{\bA'}^{\perp}(\bPhi)\hat{\bU}\hat{\bU}^{H}\right)\,,
\end{equation}
\noindent where $\bP_{\bA}^{\perp}(\bPhi) = \bI - \bP_{\bA}$ and $\bP_{\bA} = \bA\bA^{\dagger}$ is a projection onto the column span of $\bA$. The resulting SNLS problem has a reduced dimension of the parameter space and it is better conditioned than the original problem (\ref{eq:sub_fit_prob1}). A solution to this problem can be efficiently found using iterative methods such as Levenberg-Marquardt (LM) or variable projection \cite{golub2003separable}. However, these methods require initialization which can be obtained by applying the multiresolution (MR) frequency estimation algorithm \cite{kazaz2019multiresolution} on samples collected from two chirp sequences. 

\subsection{Ambiguity resolving and DDM compensation}
Due to the frequency modulation introduced by DDM MIMO scheme $K_{Tx}$ phase shifts replicas corresponding to the same target are estimated. Since DDM frequency shifts are deterministic and know, following the model (\ref{eq:model2}) we can compensate these shifts in $\bPhi$ by phase rotation of $K_{Tx}$ replica estimates as
\begin{equation} 
	\label{eq:phase_comb} 
	\Phi_{p} = \Phi_{1,p} + \Phi_{2,p} e^{-j\frac{2\pi}{K_{Tx}}} + \dots + \Phi_{K_{Tx},p} e^{-j\frac{2\pi(K_{Tx}-1)}{K_{Tx}}}\,
\end{equation} 
\noindent $p = 1, \dots, P$. From estimated phase shifts $\Phi_{p}$, $p = 1, \dots, P$, Doppler frequencies $f_{d,p}$ and velocities of the targets directly follow. However, due to large chirp repetition intervals $T_{ri}$, integer $2\pi$ wrapping occurs in the phase estimates resulting in ambiguous velocity estimation when single chirp sequence is used. This has motivated the use of multiple chirp sequences for velocity ambiguity resolving.

In traditional velocity estimation methods based on multiple chirp sequences, the phase estimates are obtained for each chirp sequence separately \cite{kronauge2010radar}. After that $2\pi$ integer unfolding of phase estimates obtained for multiple chirp sequences is performed and a later search for the minimum difference between unfolded phase estimates is done. The integer unfolding that provides the minimum difference between unfolded phase estimates is then selected as the solution to the ambiguity problem. However, this solution is sub-optimal and often works only for medium to high SNR scenarios. 

On the other hand, the solution given in (\ref{eq:sub_fit_est1}) estimates $\bPhi$ by using data from all the chirp sequences at the same time. Now, assuming that chirp repetition intervals $T_{ri}$ and shifts between the chirp sequences $T_l$, $l = 1, \dots, L$ allow for ambiguity resolving, finding the solution to the problem (\ref{eq:sub_fit_est1}) will directly result in phase estimates $\bPhi$ without integer wrapping. This will further lead to unambiguous velocity estimation. 

\begin{figure}[tp]
	\centering
	\includegraphics[width=0.45\textwidth]{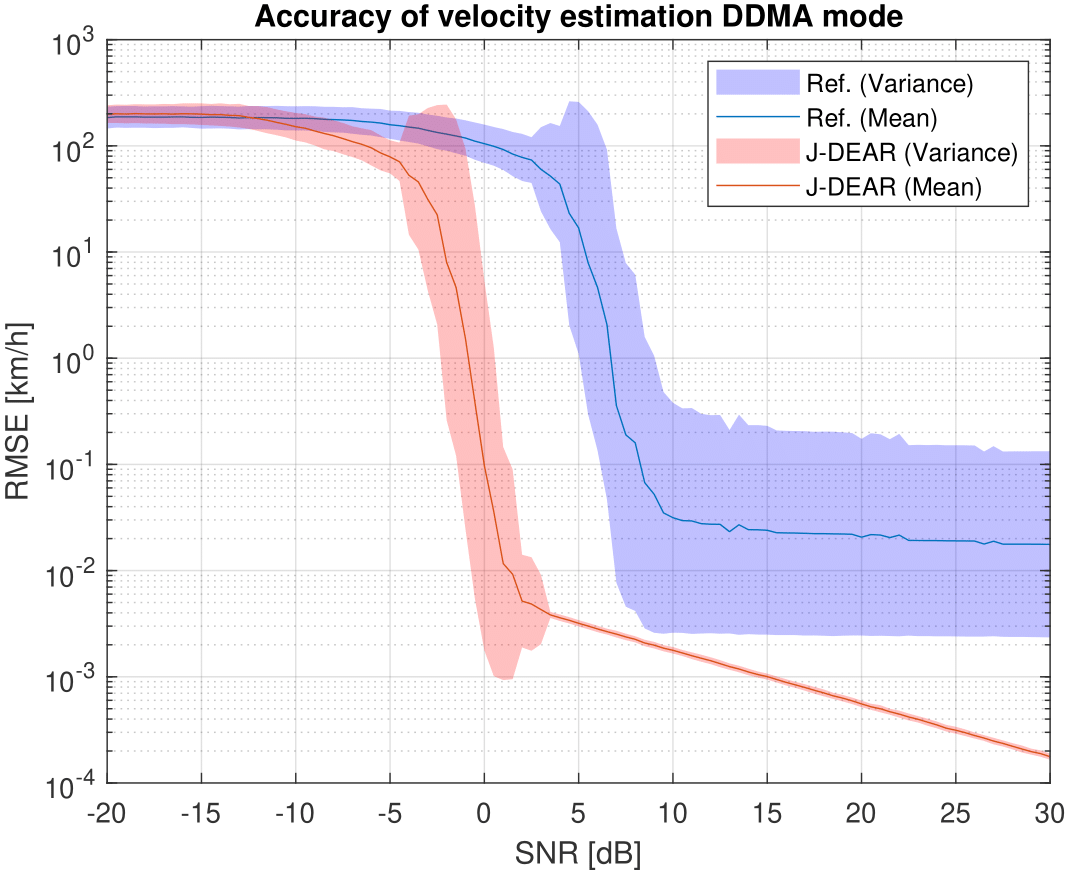}
	\caption{Root mean square error (RMSE) and its variance (shaded regions) vs. SNR for velocity estimation.}\label{fig:accuracy}
\end{figure}

\begin{figure*}[h!]
	\centering
	\includegraphics[width=0.9\textwidth]{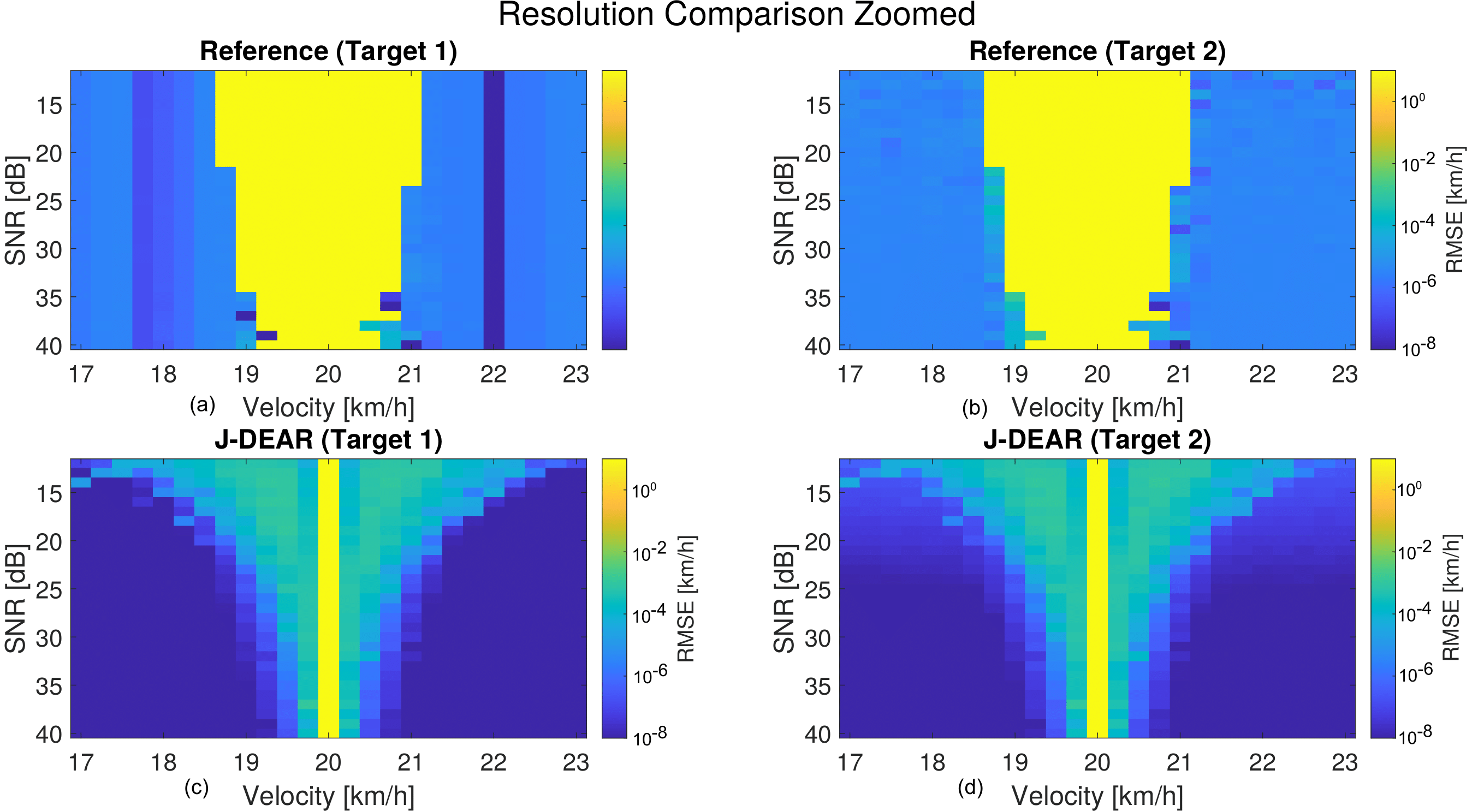}
	\caption{Root mean square error (RMSE) vs. separation and SNR of targets. RMSEs for the reference method are represented in figures (a) and (b) for 1st and 2nd target, respectively. Likewise, figures (c) and (d) represent RMSEs for J-DEAR algorithm for 1st and 2nd target, respectively} \label{fig:res_error}
\end{figure*}

\section{Results}\label{sec:results}
In this section, the numerical simulations and experimental validation of the algorithm are presented. The results show clear improvements in the accuracy and resolution of the Doppler estimation compared to a reference method \cite{kronauge2010radar}. The reference method uses traditional signal processing techniques based on FFT, phase interferometry and time shifted chirp sequences.

\subsection{Numerical simulations}
To benchmark the performance of the proposed algorithm, extensive numerical Monte Carlo simulations have been run. We consider a scenario where the radar uses two chirp sequences for velocity estimation, has $K_{Tx} = 4$, and $K_{Rx} = 4$ antennas, and uses DDM to orthogonalize Tx antennas. Each chirp sequence consists of $256$ chirps, where the $T_{ri}$ = 65.1$\mu s$ and $T_2$ = 34$\mu s$.

\subsubsection{Accuracy}

To assess the accuracy of estimation we calculate the root mean square error (RMSE) of velocity estimation for a single target in white gaussian noise. A 2D sweep of the SNR and target velocities is performed where the parameters are swept between $-20$dB to $30$dB and $-300$km/h and $150$km/h respectively. These calculations are based on $1000$ iterations for each sweep step of SNR and velocity. To ensure a fair comparison, both algorithms are receiving the same realizations of samples per iteration. The results of these simulations are shown in Figure~\ref{fig:accuracy} from which we can make the following observations: 
\begin{enumerate*}[label=\itshape\alph*\upshape)]
	\item \textbf{Coherency Gain}: If we consider that the operational regime of the algorithms is the region where RMSE is below $10^{-1}$ km/h, J-DEAR provides accurate results starting from 0 dB while the reference method reaches the same accuracy starting from 8 dB which indicates a gain of approximately 8 dB due to nearly-full coherency of processing across two chirp sequences (3 dB) and from the combining the DDM estimates (6 dB).
	\item \textbf{Grid Bias}: The average error is lower for higher SNR values for both methods. However, the error in the reference method is saturated around $10^{-2}$ km/h due to grid bias.
	\item \textbf{Doppler Dependency}: The standard deviation of RMSEs with respect to sweep velocities (i.e., shaded regions on the figure) are very different for the two methods. The performance of the J-DEAR algorithm does not depend significantly on the velocity of the target unlike the reference method whose performance heavily depends on the velocity of the target. The large variation in the standard deviation of RMSE of the reference method is caused by its ambiguity resolving where data from multiple chirp sequences is estimated interdependently. Differently, in J-DEAR algorithm estimation and ambiguity resolving are done jointly from all collected data.
\end{enumerate*}

\subsubsection{Resolution}

In this section we assess the resolution performance of the algorithms. Similar as in previous scenario, we perform 2D sweep of velocity separation between targets and SNR. We sweep SNR of the targets from $-25$dB to $25$dB while for each iteration the Doppler frequency of one target is randomly chosen within the operating interval of $-300$km/h and $150$km/h and the Doppler frequency of the second target is fixed at 4 km/h. The results are shown in figure~\ref{fig:res_error}. The following observations can be made from the figures:
\begin{enumerate*}[label=\itshape\alph*\upshape)]
 	\item \textbf{Lower Error}: From figure~\ref{fig:res_error}, by comparing the upper (reference method) sub-figures with the lower ones (J-DEAR) we can notice that the overall RMSE of J-DEAR algorithm for both targets is lower than for the reference method. 
 	\item \textbf{Improved Resolution}: Figure~\ref{fig:res_error} shows that there is a significantly increase in the resolution capabilities provided by the proposed method against the reference which can be seen the width of the yellow strip along the middle in the center two sub-figures.
\end{enumerate*}

\begin{figure}[htbp]
	\centering
	\includegraphics[width=0.3\textwidth]{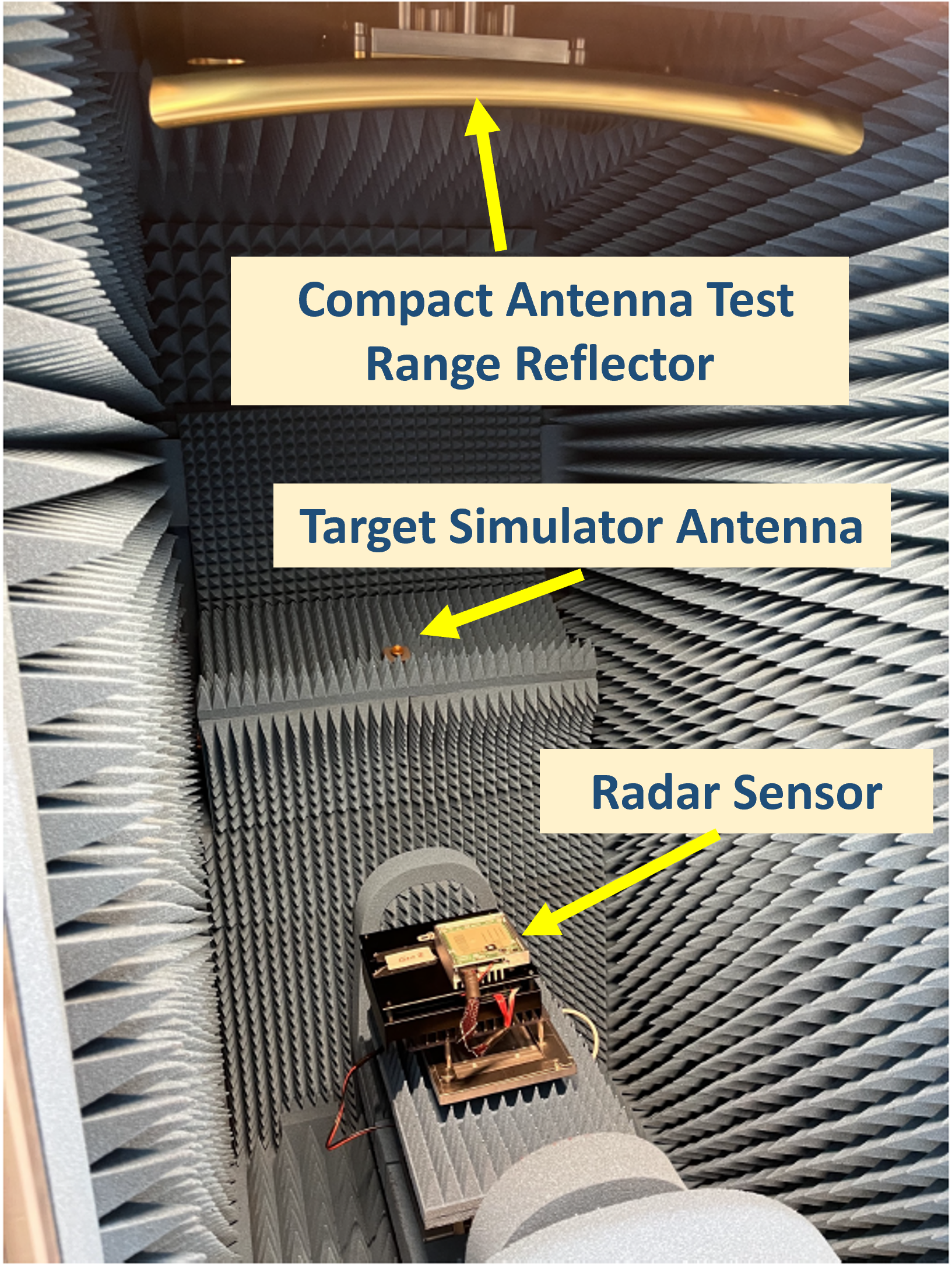}
	\caption{Experimental setup in anechoic chamber.}\label{fig:setup}
\end{figure}

\begin{figure}[htbp]
	\centering
	\includegraphics[width=0.49\textwidth]{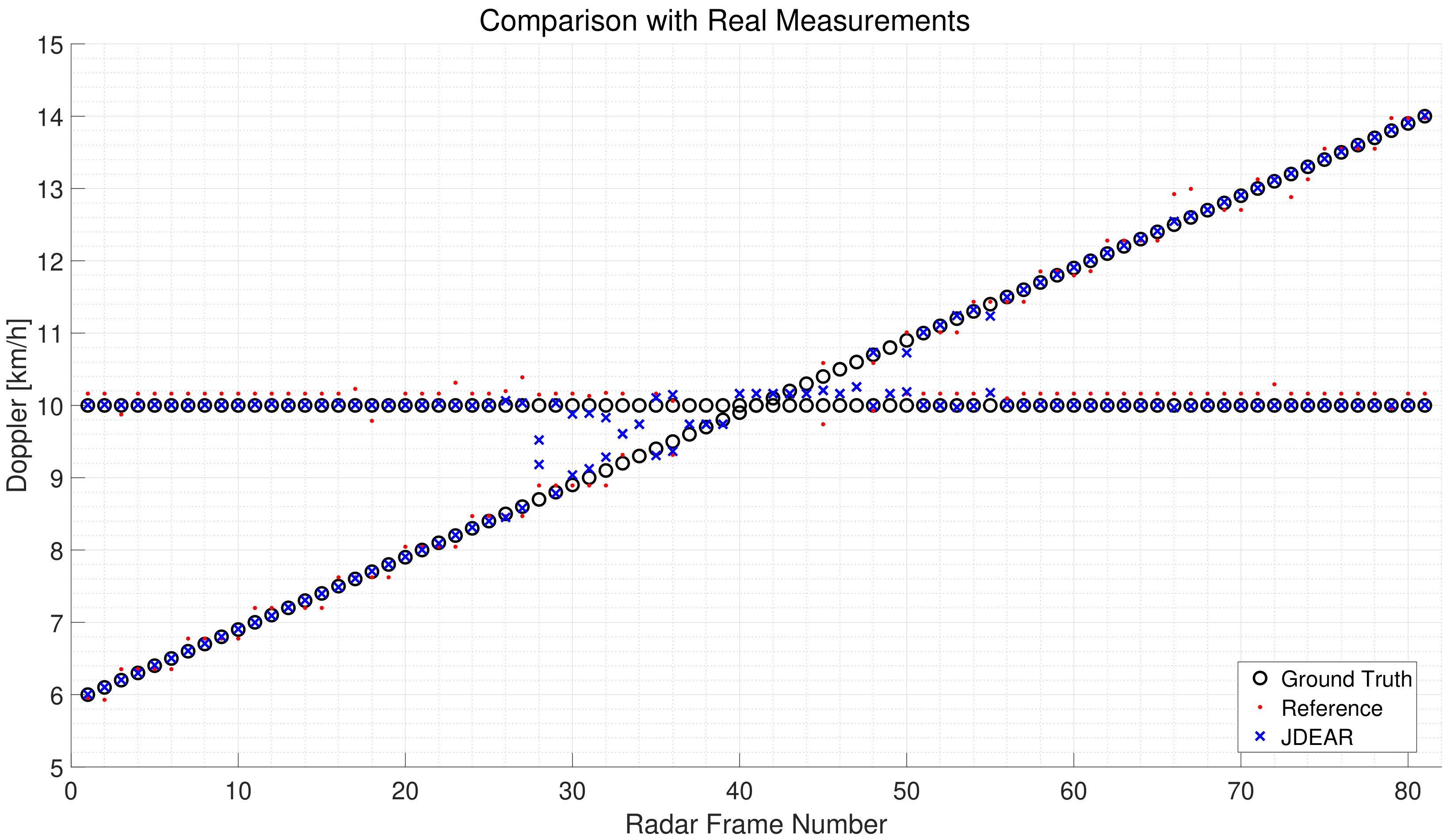}
	\caption{Results of experiments: Comparison of velocity estimates with reference method, J-DEAR and ground truth.}\label{fig:real_data}
\end{figure}
\subsection{Experiments}
To validate the results of the test, the algorithm was tested on data generated by an radar based on the NXP SAF85XX ICs. The setup is shown in figure~\ref{fig:setup} where a radar configured using the same settings as mentioned in the numerical simulations section is placed in a anechoic chamber with a target simulator attached to it. The test is a single sweep of two targets placed at $80$m from the Radar with an RCS value of $-10$dBsm. One target has a constant doppler of $10$km/h while the second target sweeps its velocity from $6$km/h and $14$km/h with a sweep step size of $0.1$km/h. The test is repeated for targets with High SNR and Low SNR and the results are shown in figure. We can clearly seem more consistent and accurate estimates using the proposed method compared to the reference.

\section{Conclusions}
In this paper, we have considered the problem of joint velocity estimation and ambiguity resolving from multiple FMCW chirp sequences in automotive radar scenarios. We have derived detailed signal models which show that range bin samples acquired using multiple chirp sequences can be transformed in data structures that have multiple shift invariance properties. Starting from this we have proposed an algorithm based on subspace estimation and nonlinear least squares optimization that exploits these structures to jointly estimate the velocities of the targets and resolves ambiguities. Furthermore, we have presented a method for DDM compensation and combining of target replicas introduced by it. 

Finally, we have presented the results of numerical simulation and experiments with radar sensors. These results show that the proposed method improves processing gain by approximately 8 dB compared to the traditional FFT-based method and increases the resolution of velocity estimation.

\bibliographystyle{IEEEtran}
\bibliography{bibliography}

\begin{thebibliography}{10}
\providecommand{\url}[1]{#1}
\csname url@samestyle\endcsname
\providecommand{\newblock}{\relax}
\providecommand{\bibinfo}[2]{#2}
\providecommand{\BIBentrySTDinterwordspacing}{\spaceskip=0pt\relax}
\providecommand{\BIBentryALTinterwordstretchfactor}{4}
\providecommand{\BIBentryALTinterwordspacing}{\spaceskip=\fontdimen2\font plus
\BIBentryALTinterwordstretchfactor\fontdimen3\font minus
  \fontdimen4\font\relax}
\providecommand{\BIBforeignlanguage}[2]{{%
\expandafter\ifx\csname l@#1\endcsname\relax
\typeout{** WARNING: IEEEtran.bst: No hyphenation pattern has been}%
\typeout{** loaded for the language `#1'. Using the pattern for}%
\typeout{** the default language instead.}%
\else
\language=\csname l@#1\endcsname
\fi
#2}}
\providecommand{\BIBdecl}{\relax}
\BIBdecl

\bibitem{patole2017automotive}
S.~M. Patole, M.~Torlak, D.~Wang, and M.~Ali, ``Automotive radars: A review of
  signal processing techniques,'' \emph{IEEE Signal Processing Magazine},
  vol.~34, no.~2, pp. 22--35, 2017.

\bibitem{8878141}
P.~van~der Meulen, P.~Kruizinga, J.~G. Bosch, and G.~Leus, ``Coding mask design
  for single sensor ultrasound imaging,'' \emph{IEEE Transactions on
  Computational Imaging}, vol.~6, pp. 358--373, 2020.

\bibitem{4291873}
S.~van~der Tol, B.~D. Jeffs, and A.-J. van~der Veen, ``Self-calibration for the
  lofar radio astronomical array,'' \emph{IEEE Transactions on Signal
  Processing}, vol.~55, no.~9, pp. 4497--4510, 2007.

\bibitem{kazaz2022multiband}
T.~Kazaz, ``Multiband channel estimation for precise localization in wireless
  networks: Algorithms, simulations and experiments,'' 2022.

\bibitem{herman2009high}
M.~A. Herman and T.~Strohmer, ``High-resolution radar via compressed sensing,''
  \emph{IEEE transactions on signal processing}, vol.~57, no.~6, pp.
  2275--2284, 2009.

\bibitem{trucco2008devising}
A.~Trucco, M.~Palmese, and S.~Repetto, ``Devising an affordable sonar system
  for underwater 3-d vision,'' \emph{IEEE transactions on instrumentation and
  measurement}, vol.~57, no.~10, pp. 2348--2354, 2008.

\bibitem{engels2017advances}
F.~Engels, P.~Heidenreich, A.~M. Zoubir, F.~K. Jondral, and M.~Wintermantel,
  ``Advances in automotive radar: A framework on computationally efficient
  high-resolution frequency estimation,'' \emph{IEEE Signal Processing
  Magazine}, vol.~34, no.~2, pp. 36--46, 2017.

\bibitem{sun2020mimo}
S.~Sun, A.~P. Petropulu, and H.~V. Poor, ``Mimo radar for advanced
  driver-assistance systems and autonomous driving: Advantages and
  challenges,'' \emph{IEEE Signal Processing Magazine}, vol.~37, no.~4, pp.
  98--117, 2020.

\bibitem{sun2022enhancing}
Y.~Sun, M.~Bauduin, and A.~Bourdoux, ``Enhancing unambiguous velocity in
  doppler-division multiplexing mimo radar,'' in \emph{2021 18th European Radar
  Conference (EuRAD)}.\hskip 1em plus 0.5em minus 0.4em\relax IEEE, 2022, pp.
  493--496.

\bibitem{jansen2019automotive}
F.~Jansen, ``Automotive radar doppler division mimo with velocity ambiguity
  resolving capabilities,'' in \emph{2019 16th European Radar Conference
  (EuRAD)}.\hskip 1em plus 0.5em minus 0.4em\relax IEEE, 2019, pp. 245--248.

\bibitem{kronauge2010radar}
M.~Kronauge, C.~Schroeder, and H.~Rohling, ``Radar target detection and doppler
  ambiguity resolution,'' in \emph{11-th International Radar Symposium}.\hskip
  1em plus 0.5em minus 0.4em\relax IEEE, 2010, pp. 1--4.

\bibitem{swindlehurst1992multiple}
A.~L. Swindlehurst, B.~Ottersten, R.~Roy, and T.~Kailath, ``Multiple invariance
  esprit,'' \emph{IEEE Transactions on Signal Processing}, vol.~40, no.~4, pp.
  867--881, 1992.

\bibitem{237536}
A.-J. Van Der~Veen, E.~Deprettere, and A.~Swindlehurst, ``Subspace-based signal
  analysis using singular value decomposition,'' \emph{Proceedings of the
  IEEE}, vol.~81, no.~9, pp. 1277--1308, 1993.

\bibitem{9555252}
T.~Kazaz, G.~J.~M. Janssen, J.~Romme, and A.-J. van~der Veen, ``Delay
  estimation for ranging and localization using multiband channel state
  information,'' \emph{IEEE Transactions on Wireless Communications}, vol.~21,
  no.~4, pp. 2591--2607, 2022.

\bibitem{viberg1991sensor}
M.~Viberg and B.~Ottersten, ``Sensor array processing based on subspace
  fitting,'' \emph{IEEE Transactions on signal processing}, vol.~39, no.~5, pp.
  1110--1121, 1991.

\bibitem{golub2003separable}
G.~Golub and V.~Pereyra, ``{Separable nonlinear least squares: the variable
  projection method and its applications},'' \emph{Inverse problems}, vol.~19,
  no.~2, p.~R1, 2003.

\bibitem{nxp}
\BIBentryALTinterwordspacing
High performance 77{GHz} rfcmos automotive radar one-chip soc. [Online].
  Available:
  \url{https://www.nxp.com/products/radio-frequency/radar-transceivers-and-socs/high-performance-77ghz-rfcmos-automotive-radar-one-chip-soc:SAF85XX}
\BIBentrySTDinterwordspacing

\bibitem{van2004optimum}
H.~L. Van~Trees, \emph{{Optimum array processing: Part IV of detection,
  estimation, and modulation theory}}.\hskip 1em plus 0.5em minus 0.4em\relax
  John Wiley \& Sons, 2004.

\bibitem{bazzi2016detection}
A.~Bazzi, D.~T. Slock, and L.~Meilhac, ``Detection of the number of
  superimposed signals using modified mdl criterion: A random matrix
  approach,'' in \emph{2016 IEEE International Conference on Acoustics, Speech
  and Signal Processing (ICASSP)}.\hskip 1em plus 0.5em minus 0.4em\relax IEEE,
  2016, pp. 4593--4597.

\bibitem{kazaz2019multiresolution}
T.~Kazaz, R.~T. Rajan, G.~J. Janssen, and A.-J. van~der Veen,
  ``{Multiresolution time-of-arrival estimation from multiband radio channel
  measurements},'' in \emph{ICASSP 2019-2019 IEEE International Conference on
  Acoustics, Speech and Signal Processing (ICASSP)}.\hskip 1em plus 0.5em minus
  0.4em\relax IEEE, 2019, pp. 4395--4399.

\end{thebibliography}
\end{document}